\documentclass{aa}

\usepackage{natbib}
\bibpunct{(}{)}{;}{a}{}{,}
\usepackage{graphicx}
\usepackage{txfonts}

\usepackage{draftcopy}
\usepackage{lscape}

\newcommand{\obj}{\,{GDS18.92$-$02.7}}

\def\halpha{\mbox{H$\alpha$}}
\def\hbeta{\mbox{H$\beta$}}
\def\lya{\mbox{Ly$\alpha$}}
\def\wb{\mbox{$B_{435}$}}
\def\wv{\mbox{$V_{606}$}}
\def\wi{\mbox{$i_{775}$}}
\def\wz{\mbox{$z_{850}$}}

\def\lesssim{\mathrel{\hbox{\rlap{\hbox{\lower4pt\hbox{$\sim$}}}\hbox{$<$}}}}
\def\gtrsim{\mathrel{\hbox{\rlap{\hbox{\lower4pt\hbox{$\sim$}}}\hbox{$>$}}}}

\begin{document}
 \title{ \lya\  emitters in the GOODS-S field:}

 \subtitle{a powerful pure nebular SED with \ion{N}{iv}] emission at z$=$5.563 \thanks{Based on observations made at the European Southern
Observatory, Paranal, Chile (ESO programme 170.A-0788) {\it The Great
Observatories Origins Deep Survey: ESO Public Observations of the
SIRTF Legacy/HST Treasury/Chandra Deep Field South.}); on observations obtained with the NASA/ESA Hubble Space Telescope
obtained at the Space Telescope Science Institute, which is operated by
the Association of Universities for Research in Astronomy (AURA), Inc.; and on observations made with the Spitzer Space Telescope, which is operated by the Jet Propulsion Laboratory, California Institute of Technology under a contract with NASA.}}

 \author{Anna Raiter
        \inst{1}
        \and
        Robert A. E. Fosbury
        \inst{2}
        \and
        Hossein Teimoorinia
        \inst{1, 3}
              }

\institute{
European Southern Observatory, Karl-Schwarzschild-Strasse 2,
Garching bei M\"unchen 85748, Germany\\  \email{araiter@eso.org}
\and
ST-ECF, Karl-Schwarzschild Str. 2, Garching bei M\"unchen 85748, Germany
\and
Institute for Advanced Studies in Basic Sciences, PO Box 45195-1159, Zanjan 45195, Iran} 



\abstract
 {The Great Observatories Origins Deep Survey (GOODS) has provided us with one of the deepest 
 multi-wavelength views of the distant universe. The combination of multi-band 
 photometry and optical spectroscopy has resulted in the identification of sources whose redshifts
 extend to values in excess of six. Amongst these distant sources are \lya\ emitters whose nature must be 
 deduced by clearly identifying the different components that contribute to the measured SED.}
 {From a sample of \lya\ emitters in the GOODS-S field with uncontaminated photometry and 
 optical (red) spectroscopy, we select a spatially compact object at a 
 redshift of 5.563 (\lya) that shows a second emission line, identified as \ion{N}{iv}] 1486~\AA. The SED 
 is modelled in a way that accounts for both the \ion{N}{iv}] line emission and the photometry in a self-consistent way. }
 {The photoionization code CLOUDY is used to calculate a range of nebular models as a function 
 of stellar ionizing source temperature, ionization parameter, density and nebular metallicity. We compare the theoretical and observed magnitudes
 and search for the model parameters that also reproduce the observed \ion{N}{iv}] luminosity and equivalent width.}
 {A nebular model with a hot blackbody ionizing source of around 100~kK and a nebular 
 metallicity of $\sim$5\% of solar is able to fit the observed SED and, in particular, explain the large
 apparent Balmer break which is inferred from the pure stellar population model fitting
 conventionally applied to multi-band photometric observations. In our model, an apparent spectral break is produced by strong
 [\ion{O}{iii}]  4959, 5007~\AA\, emission falling in one of the IR bands (IRAC1 in this case). A lower limit on the total 
 baryonic  mass of a model of this type is $3.2 \times 10^8~{\rm M}_{\odot}$. }
{It is argued that objects with \lya\ emission at high redshift that show an apparent Balmer break may
have their SED dominated by nebular emission and so could possibly be identified with very young starbursting galaxies
rather than massive evolved stellar populations. Detailed studies of these emission nebul\ae\  with large telescopes will
provide a unique insight into very early chemical evolution.}


\maketitle
%
%

\section{Introduction}
The \lya\  emission line is a beacon that can be recognised from the
ground in objects with redshifts larger than about 1.8. When seen on its own,
its interpretation is complicated by radiative transfer effects both within the
emission region itself and in the intervening Lyman forest. With the detection
of other emission lines and/or a recognisable continuum, a spectrum can however
provide a rich source of information about physical conditions and chemical
composition. Of particular interest for the identification of primordial
starforming objects are the so-called `dual-emitters': sources that emit both
\lya\ and the \ion{He}{ii} line at 1640~\AA\,\citep{daw04, nag05, ouc08} which are interpreted as very metal-poor nebul\ae\  ionized by
massive, hot and presumably metal-free stars. So far, intermediate band imaging
surveys have not been very successful in discovering genuine, stellar-ionized H/\ion{He}{ii} dual emitters \citep{nag08}.

The Great Observatories Origins Deep Survey (GOODS) spectroscopic observations 
in the Chandra Deep Field South \citep{van05,van06,van08,van09,pop09} have accumulated, using the FORS2 and VIMOS
instruments on ESO's Very Large Telescope, spectroscopy of essentially all accessible sources.
The source GDS~J033218.92$-$275302.7 (hereafter \obj) was selected for the spectroscopic 
programme as a \wv-band dropout and has a blue \wi-\wz\, colour due to a 
strong \lya\ emission line falling in the \wi-band. It was subsequently studied in
the context of SED fitting by \citet{sta07} and by \citet{wik08}
who both conclude, on the basis of its apparently large Balmer break, that it is
a high mass, old stellar population. Since the restframe UV images obtained with
the ACS on HST show a very small angular size, the implication is that this is an 
example of an ultradense spheroid at a very early epoch. A significant fraction of other \lya\ emitters from 
GOODS, showing apparent Balmer breaks, have also been interpreted as evolved, massive stellar populations \citep{pen09}.

The presence of strong emission lines in the UV spectrum carries implications
about the effect of nebular emission processes on the rest of the observable SED.
There are circumstances in which the combination of nebular continuum from
bound-free and two-photon emission together with emission lines can influence
the appearance of, or even completely dominate, an underlying stellar population
throughout the UV, optical and NIR spectrum. In this paper, we exploit the
presence of a second emission line, identified as the [\ion{N}{iv}], \ion{N}{iv}]
intercombination doublet at 1483, 1486~\AA, to estimate the implied SED and
compare with the extensive GOODS photometry of this object.

The presence of significant \ion{N}{iv}] emission is rather rare in known astronomical
objects. At low redshifts it is seen in some compact planetary nebul\ae\  and in
peculiar very hot stars such as $\eta$ Carin\ae\  \citep{dav86}. It is also sometimes
detected in AGN and quasar spectra but with an intensity that is low with
respect to lines such as \ion{C}{iv} and \ion{N}{v}, although there are rare examples of `Nitrogen Loud' quasars that
exhibit unusually strong nitrogen lines, e.g. \citet{bal03}. At a redshift of 3.4, the Lynx arc \citep[hereafter F03]{fos03}, 
is known to show \ion{N}{iv}] with a strength comparable to other UV
intercombination lines such as \ion{O}{iii}] and \ion{C}{iii}] and the resonance doublet \ion{C}{iv}. This object is
interpreted as an \ion{H}{ii} region ionized by very hot stars ($T_{\rm eff} \sim 80$~kK) with a
nebular metallicity a few per cent of solar. In this case, the entire observed SED is nebular 
in origin and the stellar emission is not detected longward of \lya. \citet{gli07} report
two \ion{N}{iv}] emitting low luminosity `quasars' one of which, DLS~1053$-$0528 at $z=4.02$, 
has an intensity ratio of \ion{N}{iv}]/ \ion{C}{iv} $=1.9$, a weak continuum and an SED which is considerably 
bluer than a typical quasar. This object has relatively narrow lines and may be more closely related to  
the Lynx arc than to an AGN.

In this paper we examine a sample of \lya\ emitters from the GOODS spectroscopic 
data with redshifts spanning the range $3.5\leqslant z \leqslant 5.9$ and present
new photometric measurements from the most recent releases of the HST, VLT and Spitzer
imaging data. For our selected \ion{N}{iv}] emitter, we use a photoionization modelling code to investigate 
the range of model nebul\ae\ \ that will produce the observed \ion{N}{iv}] emission flux. 
Given the GOODS broad-band photometric measurements that extend from the U-band 
to the Spitzer MIPS $24~\mu$m channel, we calculate the contribution of the nebular line and 
continuum emission to these bands in order to constrain the nature of any non-nebular light
that may contribute. In addition to modelling the SED of this one object, we address the possibility
that nebular emission may contribute significantly to other sample members, 
a possibility that has already been suggested by \citet{zac08}.
Our conclusions are of relevance to the interpretation of the SEDs of other
high redshift \lya\ emitters, where the emission line strength suggests that the nebula 
may be a significant contributor to the spectrum longward of the \lya\ line. We adopt a cosmology with $\Omega_\Lambda = 0.7, \Omega_{\rm M}=0.3$ and $ {\rm}H_0 = 70$~km~s$^{-1}$~Mpc$^{-1}$. 
The $AB$ magnitude system is used throughout.

\begin{figure}[h!]
\centering
\includegraphics[width=8.8cm]{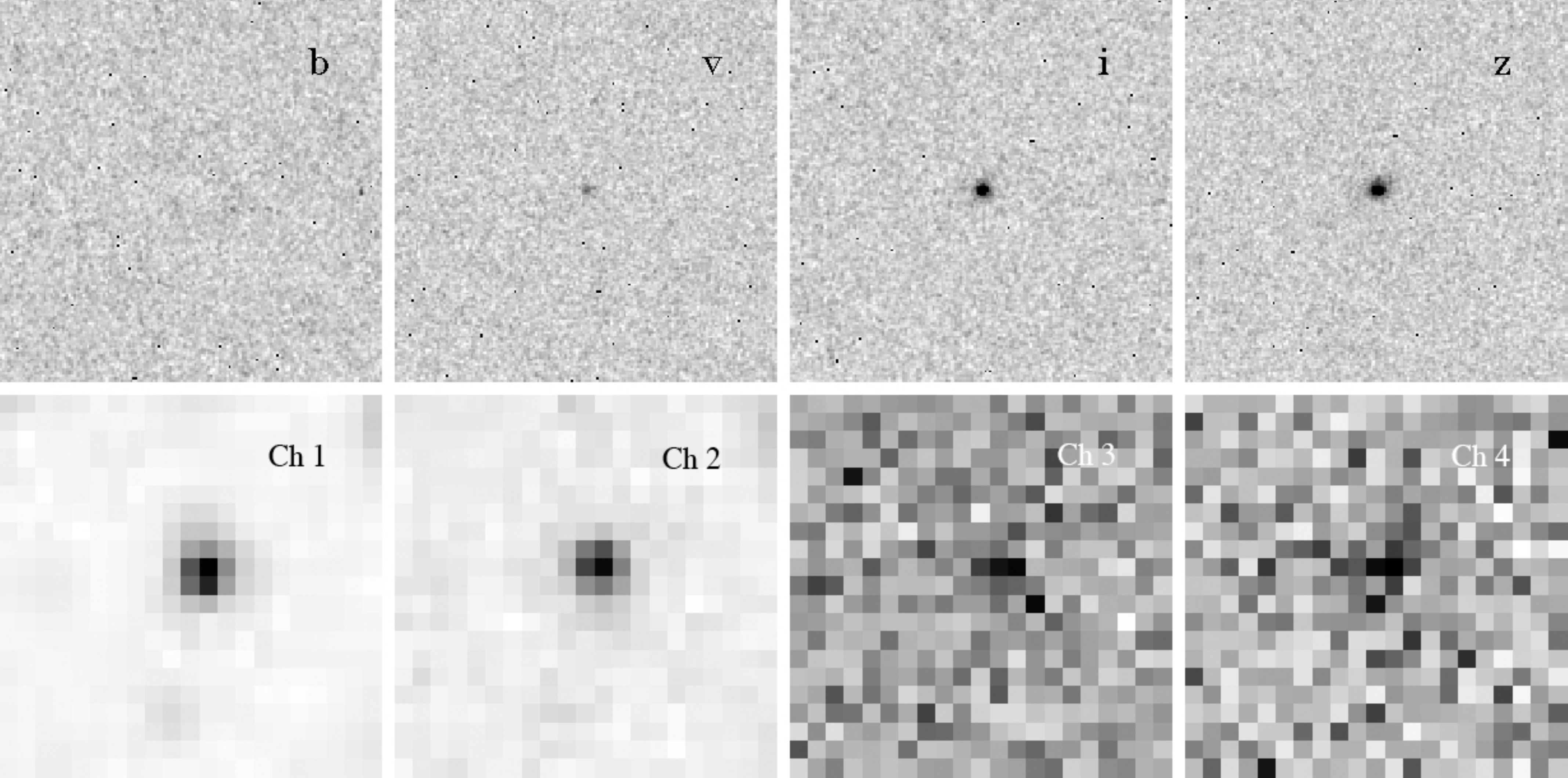}
\caption{Image cutouts of \obj\ from HST/ACS (top-row) and Spitzer/IRAC (bottom-row). The ACS stamps are 5~arcsec on a side and the IRAC stamps are 25.6~arcsec.}
\label{cutouts}
\end{figure}

\section{Data}

\subsection{Sample}

We have spectroscopically selected a sample of \lya\ emitters from the publicly available data for the 
southern field of the Great Observatories Origin Deep Survey
\citep[see][for a review of 
the GOODS project]{dic01,gia04}. The multi-wavelength observations provide us with information 
on the $U$ (VLT), \wb\, \wv\, \wi\ and \wz\ (HST), $Js, H, Ks$
(VLT), 3.6, 4.5, 5.8, 8.0 and 24~$\mu m$ (Spitzer) bands. Many other 
observations have been carried out, including deep integrations
in the X-ray and radio domains from Chandra and the Very Large 
Array, respectively.

The spectroscopic observations carried out with the ESO VLT FORS2 
instrument yield a wavelength coverage from approximately
0.55--1~$\mu m$ with a resolving power of R~$\approx$~660. 
This programme has produced about one thousand 
redshift determinations between redshift 0.5 and 6.2, 
with more than one hundred Lyman break galaxies confirmed at redshifts
beyond 3.5 \citep{van05,van06,van08}.

From the GOODS data product archive {\tt http://www.eso.org/science/goods},
we used the FORS2 online engine to select all the qf (quality flag) $=$ A `emission' objects
at $z \geqslant 3.5$. A few objects were excluded because of spectral 
contamination, one because it showed no \lya\ emission and one
because of positional uncertainty. This process yielded 23 objects from 
which an additional five were excluded due to large photometric errors in some passbands.

\subsection{Photometry}

Due to the different telescopes in use, the GOODS images are characterised by 
a large range in point spread function (PSF) dimension. An unbiased
estimate of the colours is essential for analysis of the
SED. We have used a PSF-matched photometric technique to obtain the $AB$ 
magnitudes of each object in each band in a given aperture size. In the presence 
of potential contamination by neighbouring objects, we have used an
optimum circular aperture which maximises the S/N of each object
in all passbands simultaneously. For example, in the case of the match between
ACS and groundbased $Ks$ or IRAC images, components may be 
clearly resolved by ACS but be merged in IRAC, or very red
objects may be detected at the longer wavelengths but show no counterpart in the optical
images. A curve-of-growth analysis yielded the optimum uncontaminated aperture.

Each source was examined in a $12 \times 12$~arcsec$^2$ cutout and a Monte-Carlo 
method used to estimate the background. 
The photometric measurements were made with Sextractor \citep{ber96},
where the background used was consistent with Monte-Carlo simulations. Finally, by
selecting some bright but unsaturated stars near each galaxy, we computed the
convolution kernel required to match the PSF of all the images
to that with the largest. We generated
about 200 PSF kernels of stars in different filters in order to find the best kernel using
a $\chi^2$ minimisation method. All magnitudes were corrected to a 
6~arcsec circular aperture. The results for all bands appear in Table~\ref{photometry}.
We also use Sextractor to measure the `half-light' (or `effective') radius of each galaxy in the \wz-band (also listed in Table~\ref{photometry}).


\subsection{The {\rm  \ion{N}{iv}]} emitter}

\begin{figure}[h!]
\centering
\includegraphics[width=\columnwidth]{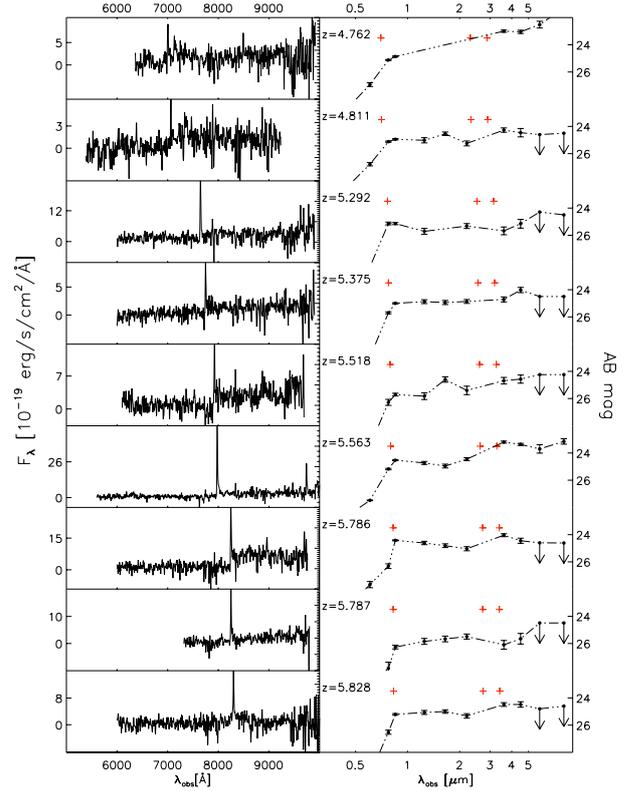}
\caption{FORS2 spectra and photometric SED -- part 2 (redshifts 4.8--5.9).}
\label{sed2}
\end{figure}

\begin{figure}[h!]
\centering
\includegraphics[width=\columnwidth]{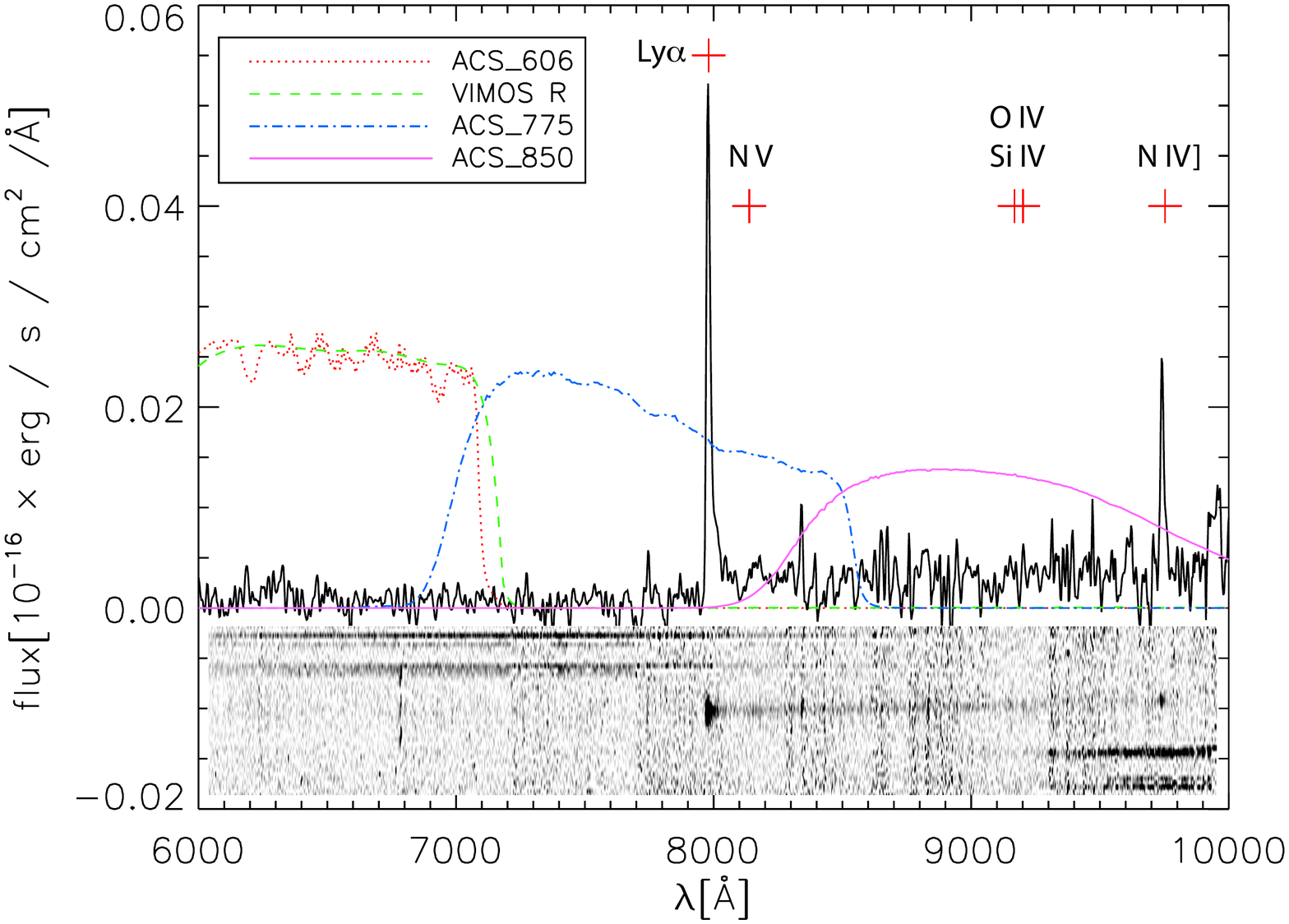}
\caption{FORS2 spectrum of source \obj\ showing 
the \lya\ and \ion{N}{iv}] emission lines and the ACS and VIMOS filter passbands. 
The expected positions of the \ion{N}{v}, \ion{O}{iv} and \ion{Si}{iv} lines are also marked.}
\label{spec_detail}
\end{figure}

Of the 18 \lya\ emitters in our sample, only one shows a clearly detected second emission line. Galaxy  \obj\, with \wz$=$24.5,
was selected as a \wv--band dropout and has been confirmed to be 
at redshift 5.563 (redshift of the \lya\ line uncorrected for absorption). The spectrum, included in 
Figure \ref{sed2} and shown in detail in Figure~\ref{spec_detail}, was extracted from several individual
FORS2 multi-object masks and corresponds to a total exposure time of 14.4~ksec 
\citep{van06}. Detailed measurements of the spectrum of this object are 
presented in \citet{van09b} and the parameters relevant to our modelling 
are summarised in Table~\ref{physprop}.

\begin{figure}[h!]
\centering
\includegraphics[width=\columnwidth]{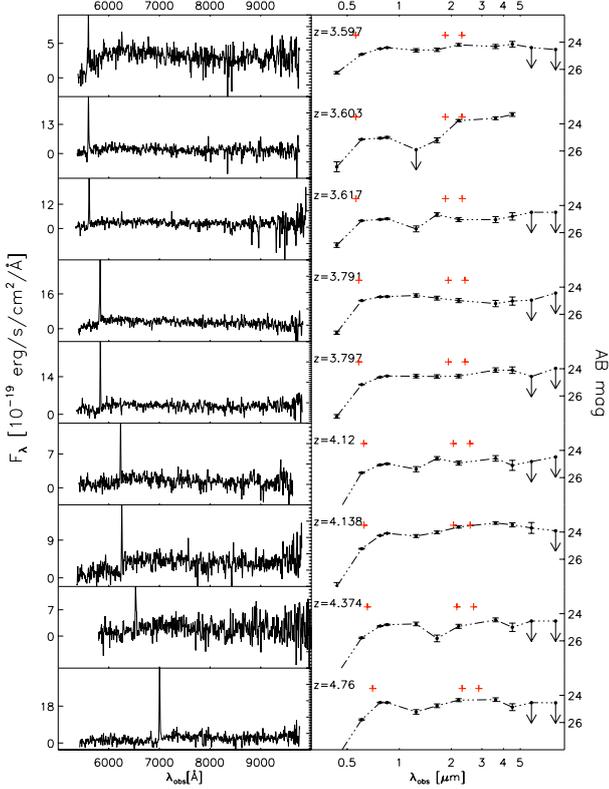}
\caption{FORS2 spectra and photometric SED -- part 1 (redshifts 3.6--4.8). The photometric points are: 
\wb\, \wv\, \wi\ and \wz\ (HST), $Js, H, Ks$
(VLT), 3.6, 4.5, 5.8, 8.0 (Spitzer).
The sources are presented in order of increasing redshift and so 
can be cross-referenced to Table~\ref{photometry}. The (red) + `plusses' 
for each SED represent the wavelengths of \lya, the 4000~\AA\ Balmer 
break and the [\ion{O}{iii}] 5007~\AA\ line. The photometric error bars are 1$\sigma$.}
\label{sed1}
\end{figure}

\begin{table}
\caption{Summary of the physical quantities derived from the optical 
spectral features of \obj\ taken from \citet{van09b}.}
\label{physprop} 
\centering 
\begin{tabular}{l l l} 
\hline\hline 
Quantity & Value & Comment\\ 
\hline 
$z$ ( \lya\ peak)     & 5.563 & no abs. correction\\ 
$L$( \lya\ )         & 3.8$ \pm 0.3 \times$10$^{43}$ erg~sec$^{-1}$\\
${\rm EW}_{0}$( \lya\ )   & 59$^{+195}_{-29} $ ~\AA~(restframe) & 89~\AA\ with Ctm$_{\rm phot}$\\
$z$ (\ion{N}{iv}])      & 5.553 (1486.5~\AA) & high-density limit\\
$L$(\ion{N}{iv}])       & 1.3$^{+0.3}_{-0.4}  \times$10$^{43}$ erg~sec$^{-1}$ &\\
${\rm EW}_{0}$(\ion{N}{iv}])& 22$^{+64}_{-10} $~\AA~(restframe) & 33~\AA\ with Ctm$_{\rm phot}$\\
$L$(\ion{Si}{iv})       & $\leqslant 0.11\pm0.07 \times10^{43}$erg~sec$^{-1}$ &(1393.8, 1402.8~\AA)\\
$L$(\ion{N}{v})       & $\leqslant 0.11\pm0.06 \times10^{43}$erg~sec$^{-1}$ &(1238.8, 1242.8~\AA)\\
\hline \end{tabular}
\begin{list}{}{}
\item[] Equivalent width values are also given using the continuum level derived from the photometry (Ctm$_{\rm phot}$).
\end{list}
\end{table}

The main spectral features are the \lya\ emission line (restframe equivalent width (EW) $\sim$60~\AA),
the break of the continuum just blueward of
the line, and the intercombination
emission doublet [\ion{N}{iv}], \ion{N}{iv}]~1483.3, 1486.5~\AA, whose intensity ratio can be 
used to estimate electron density
(see Figure~\ref{spec_detail}). There is no clear detection of \ion{Si}{iv} 1394, 
1403~\AA~in absorption or emission nor of \ion{N}{v} 1239, 1243~\AA\ in emission. 
\citet{van09b} also note that the object is extremely compact but marginally resolved in the ACS images.

\section{SED modelling}

In this section, we first discuss the modelling of  \obj. 
We then use a grid of similar photionization models in order to assess the 
potential importance of nebular emission for interpreting the SED of the other 
sources in our sample.

\subsection{GDS~J033218.92-275302.7}

Two previous studies have attempted to fit the SED of  \obj\, using the
apparently large Balmer break to resolve the age-extinction degeneracy. \citet{sta07}, 
using ID$=$32\_8020, and \citet{wik08}, as BBG 5197, both obtain an age of 0.9~Gyr, a stellar 
mass of close to $10^{11}{\rm M}_\odot$\ and zero extinction. The latter authors, who use the 
two longer wavelength IRAC channels, 5.7 and 8.0~$\mu$m, note that their fit is worsened by a deviant 5.7~$\mu$m
point.

Our nebular modelling is motivated by the realisation that a source with such strong emission lines in the FUV
will necessarily have nebular contributions at longer wavelengths. Also, by analogy with the Lynx arc at the lower redshift of 3.4, 
it is possible that the significant jump in brightness between the K-band and the IRAC 3.6~$\mu$m filter could be caused by strong [\ion{O}{iii}] 
4959, 5007~\AA\, emission lines producing an apparent, but false Balmer break signal. The nature of the photoionizing source can be inferred from the strength of \ion{N}{iv}] 
and the non-detection of \ion{N}{v} which also lies in a favourable 
region of the FORS2 spectrum. Again, by analogy with the arguments already used for the Lynx arc \citep[F03;][]{bin03,vil04}, 
the limit on this line ratio can be used to argue for a hot blackbody-like source rather than the hard power law spectrum associated with an AGN. The absence of detectable radio \citep{kel08}\ or 
X-ray \citep{gia02}\ emission from the source at very low flux levels is also evidence against an AGN interpretation. 

Since our modelling is designed 
to provide a `proof of concept' rather than a detailed SED fit, we use a set of blackbodies ranging from 60 to 120~kK as ionization sources. The use of hot stellar models having low 
or zero metallicity would have little effect on the gross properties of the predicted spectrum although they would be required to fit individual emission line fluxes if more observations 
were available. The {\em effective} temperature (the tempterature of the black body which produces the same bolometric luminosity)
of such stellar models  would be somewhat lower (by $\approx 10-15$~kK) than the blackbody temperatures we use here.

The principal requirement of a model is to produce the observed luminosity of the \ion{N}{iv}] doublet. The combination of density and ionization parameter,  $U$ (defined at the inner edge of the nebula) is constrained by the measured 
upper limit on the source size:

\begin{eqnarray}
U = \frac{Q({\rm H})}{4 \pi r_{\rm in}^2  n_{\rm H} c} \nonumber
\end{eqnarray}

\noindent where $Q$(H) is the rate of hydrogen ionizing photons radiated by the source, $r_{\rm in}$ is the inner radius of the cloud,
$n({\rm H})$ is the hydrogen number density of the cloud and $c$ is the speed of light. Since, for all of our models, the depth of the ionized gas is small compared with the 
observed size of the source, we associate $r_{\rm in}$  approximately with the PSF-corrected measurement from the \wz\ ACS image of
 $r_{\rm e}$\ (Table~\ref{photometry}) which corresponds to 580~pc. 
The observed wavelength of \ion{N}{iv}], compared with that of the blue-absorbed \lya\ line, suggests that the doublet is emitted from a region 
that is in the high density regime (\citet{van09b} show an enlarged plot of the spectrum of this region) with $n_{\rm e}\gtrsim10^{4.5}$~cm$^{-3}$, 
see: \citet{ost06}. We note, however, that in the Lynx arc, the density derived from \ion{N}{iv}] doublet is significantly higher than the value of $\lesssim3000$~cm$^{-3}$ 
derived from those of \ion{C}{iii}] 
and \ion{Si}{iii}] \citep{vil04}\footnote{But note that, for the observed line ratio,  the \ion{N}{iv}] density in this paper should be $10^5$~cm$^{-3}$, 
according to the method presented in \citet{kee95}, rather than the quoted 
2$\times 10^4$~cm$^{-3}$.}.

We use the photoionization code CLOUDY version 08.00 \citep{fer98} to calculate the models of \ion{H}{ii} regions exposed
to the blackbody radiation. We assume an ionization bounded spherical geometry (i.e., the outer radius of the cloud is not defined)
which means that the ionizing photon escape fraction is zero. The nebular gas is assumed to have constant density. 
From the output of the CLOUDY models, which includes both nebular continuum and line emission, we compute the theoretical magnitudes for all the photometric filters to compare 
with the GOODS observations. 

We have calculated a grid of models exploring the parameter space of $T_{\rm bb}$ of the blackbody, the ionization parameter ($U$)
and the nebular metallicity ($Z_{\rm neb}$),
investigating how the changes in these parameters
affect the \ion{N}{iv}] luminosity and the observed SED. We have assumed the solar relative abundances (as defined in CLOUDY) with the realisation that this is unlikely to be entirely 
appropriate at this epoch. In practice, this means that our calculated magnitudes will be affected by the abundance-dependent strength of lines from individual ions. The most 
important of these is the [\ion{O}{iii}]~4959, 5007~\AA\, doublet that is included in the IRAC1 band. Note that, with the range of abundance considered here, oxygen is 
not the dominant coolant and so this emission line strength is proportional to abundance. Consequently, our principal assumption regarding abundances is that of a solar N/O ratio. 
Guided by the example of the Lynx arc, we have performed our modelling assuming that dust is not present. However,
 given the conclusions of the \lya\ profile modelling discussed by 
\citet{van09b}, we do later examine the possibility of some dust extinction being present.

In addition to their fluxes, the emission line equivalent widths (EW) are an important constraint 
because of their dependence on the strength of the continuum. In our models, the continuum under the  \ion{N}{iv}]  
line is dominated by hydrogen two-photon emission which becomes collisionally de-excited above an electron density 
of $\sim10^{4}$~cm$^{-3}$. The observed EW (Table~\ref{physprop}) provides an additional constraint on $T_{\rm bb}$, $U$,  $n_e$\ 
and $Z_{\rm neb}$.

Guided by the results of  $\chi^2$ testing of our model grid against the observed photometry and \ion{N}{iv}] emission line flux, we choose a `fiducial' model that represents our best estimate for 
the model parameters (see Figure \ref{sed_2}). Because of our spectroscopic observation, high weight is accorded to the 
luminosity and the equivalent width of the \ion{N}{iv}] doublet. This gives the parameters shown in Table~\ref{fidpam}.

\begin{figure}[h!]
\centering
\includegraphics[width=\columnwidth]{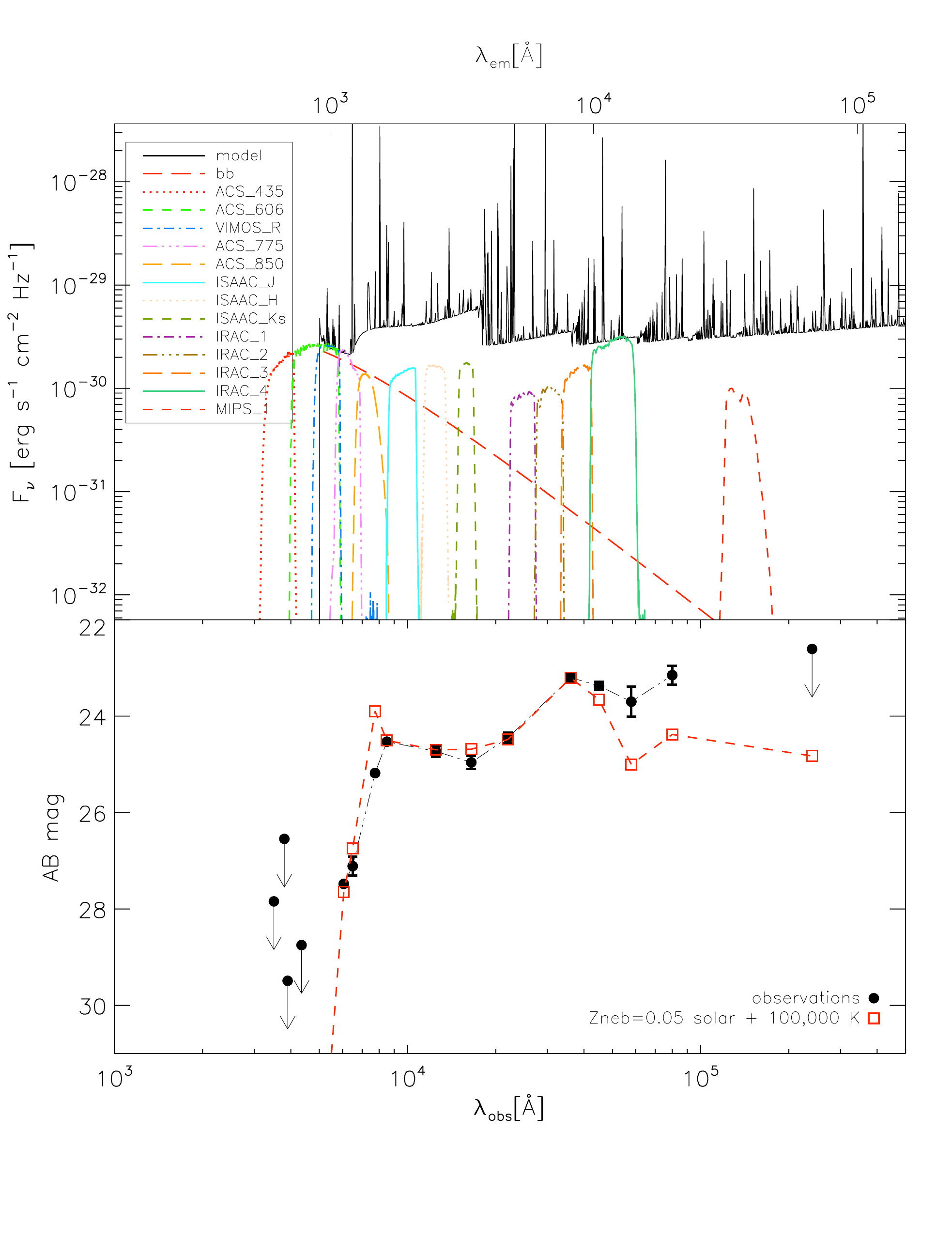}
\caption{Observed and modelled SED of  \obj. {\em Top panel:} the computed nebular spectrum from the fiducial model, the Rayleigh-Jeans tail of the ionizing blackbody 
spectrum and the GOODS filter passbands. {\em Bottom panel:} the observed and computed photometric SED. 
Additionally VIMOS $R$\ (VLT) and MIPS $24~\mu$m (Spitzer) magnitudes are included in the SED \citep{san09}.
The observed points have 1$\sigma$\ error bars or are 1$\sigma$\ upper limits. 
The points below \lya\ have been corrected for IGM absorption. Our model does not model the \lya\ radiative transfer 
and so the \wi\ point is really an upper limit. Note that the $Ks$\ band samples almost pure nebular continuum.}
\label{sed_1}
\end{figure}

\begin{figure}[h!]
\centering
\includegraphics[width=\columnwidth]{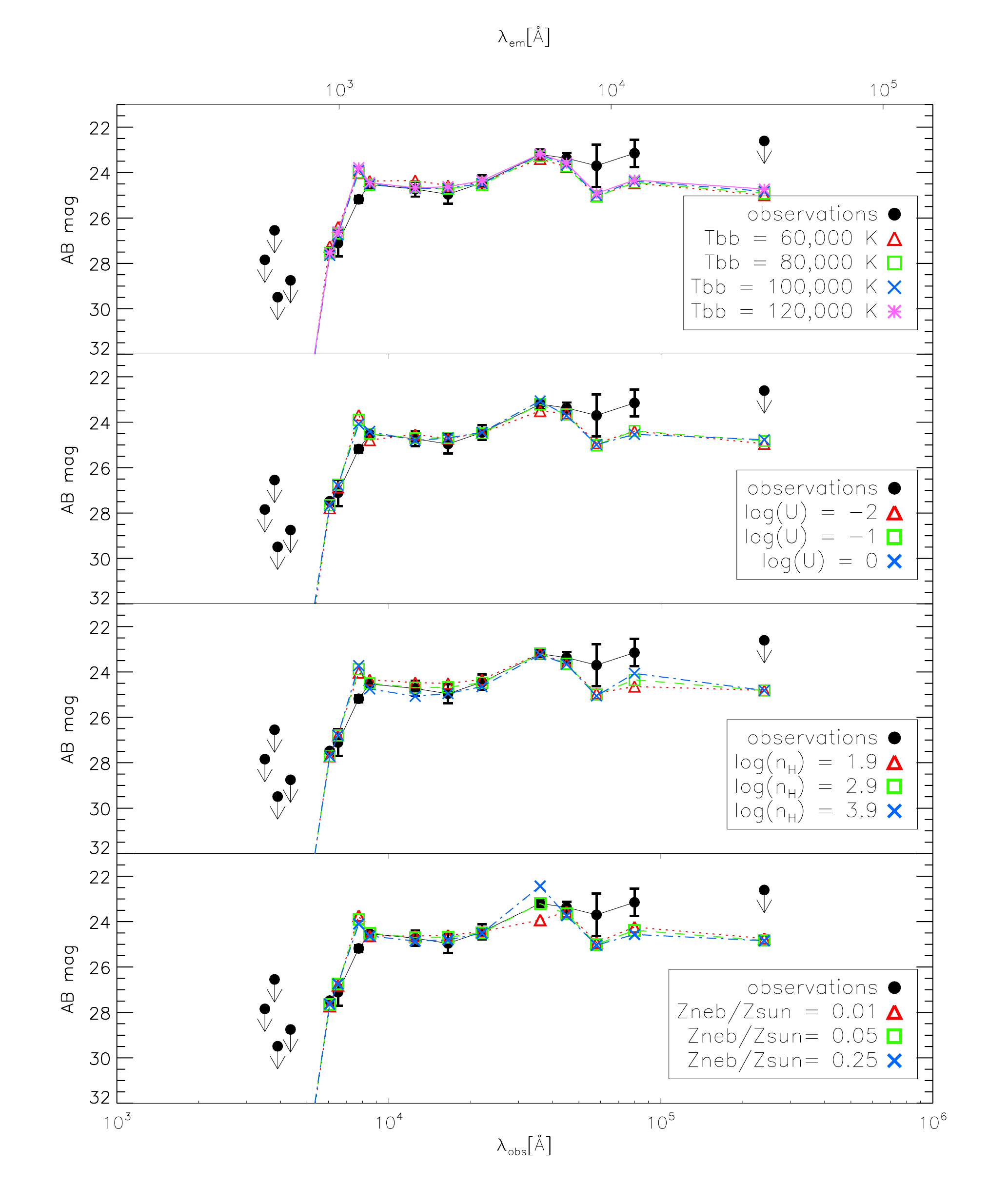}
\caption{As Figure~\ref{sed_1} but showing the effects of varying the four model parameters: {\em upper panel:} $T_{\rm bb}$,  {\em second panel:} $\log(U)$, {\em third panel:} $\log(n_H)$  and  
{\em lower panel:} $Z_{\rm neb}$. Note the effect of metallicity on the IRAC1 band  (due to [\ion{O}{iii}])\ and the effect of the density on 
IRAC4 due to changes in the \ion{He}{i} 10830\AA\ line. To improve visibility in this plot, note that the error bars (but not the upper limits) are 3$\sigma$.}
\label{sed_2}
\end{figure}

\begin{table}
\caption{Parameters of the fiducial photoionization model.}       
\label{fidpam}      
\centering                   
\begin{tabular}{l l l}       
\hline\hline                 
Parameter & Value & Unit \\    
\hline                        
$T_{\rm bb}$ &100 & kK\\
$\log(U)$ &$-1$\\
$Z_{\rm neb}$ & $0.05~Z_{\odot}$\\
$Q$(H) & $3 \times 10^{55}$& s$^{-1}$\\
$n_{\rm H}$ & $10^{2.9}$ & cm$^{-3}$\\
$r_{\rm in}$ & $10^{21}$ & cm\\
\hline                              
\end{tabular}
\end{table}

It is significant to note that this is very close to the set of parameters as derived 
for the Lynx arc (F03) based on the fitting of many individual emission lines.
There are differences in the assumed geometry and the particle density 
but these have only minor effects on the emitted spectrum. We have used a 
higher density in order to keep the nebula small enough to be consistent with 
the imaging data. The SED derived from this fiducial model is shown in 
Figure~\ref{sed_1} and the luminosities of some principal emission 
lines in Table~\ref{em_lines}. This model produces an \ion{H}{ii} region with 
an ionization structure and electron temperature distribution that  results in 
an radially-averaged electron temperature of 23~kK.

\begin{table}
\caption{Fiducial model emission line spectrum.}             
\label{em_lines}      
\centering                          
\begin{tabular}{l l l}        
\hline\hline                 
$\lambda_{rest}$ & $\log$(L [erg s$^{-1}$]) & $I/I_{\hbeta}$ \\    
\hline                        
\ion{O}{vi} 1035 & 40.984 & 0.007\\      
\ion{Si}{iii} 1207 & 41.589 & 0.026 \\
\lya\ 1215 & 44.724 & 36.011\\
\ion{N}{v} 1240 & 42.229& 0.115\\
\ion{Si}{iv} 1397    & 42.564  & 0.249\\
\ion{O}{iv} 1402 & 42.416 & 0.177 \\
\ion{N}{iv}] 1486 &42.938 & 0.590\\
\ion{C}{iv} 1549 & 44.054 & 7.698\\
\ion{He}{ii} 1640 & 43.026 & 0.7217 \\
\ion{C}{iii}] 1909    & 42.985 & 0.657 \\
$[$\ion{O}{ii}] 3727 & 41.411 & 0.018\\
\ion{He}{ii}  4686 & 42.127 & 0.0911\\
$[$\ion{O}{iii}] 5007 &43.936 & 5.865\\
\halpha\ 6563 & 43.625 & 2.867\\
\hline                                   
\end{tabular}
\begin{list}{}{}
\item[] Note that no resonance line absorption is 
accounted for in these values. For the doublets, including [\ion{O}{iii}] 4959, 5007~\AA, the number represents the sum of the two lines.
\end{list}
\end{table}

{In this model, the $Ks$ band samples mostly nebular continuum while IRAC1 is dominated by very strong [\ion{O}{iii}] line emission resulting in a jump of 1.3 magnitudes 
between restframe wavelengths of 0.33 and 0.54~$\mu$m. The IRAC2 band is significantly influenced by  \halpha\ and IRAC4 by HeI 10830~\AA, this latter line becoming
stronger at higher density (see the third panel of Figure~\ref{sed_2}). Examination of Figure~\ref{sed_2} (which has 3-$\sigma$\ error-bars) shows that our fiducial model falls below the low signal/noise data from the IRAC3 and IRAC4 filters (see Figure~\ref{cutouts}). The IRAC3 band in our model is dominated by bound-free continuum and is little affected by changing model parameters within the range we have considered. Although the IRAC4 band model also falls below the data, the amount can be lessened significantly by changing the density or adding a higher density component. Given the marginal formal significance of these two discrepancies we do not consider it justified to attempt to add an additional source that would contribute flux at these wavelengths without destroying the fit at shorter wavelengths. In order to provide enough light from stars, the population would need to be old, massive and significantly reddened. \citet{van09b} have examined SED fits using both single and fixed stellar population models. These both require an old, $>700$~Myr for a single and $\sim 400$~Myr for a mixed, massive, $\sim 10^{11}$~M$_\odot$, population in order to provide the light seen in the IRAC bands. Such an addition would completely change the nature of the young, low-mass model we are proposing.}

In addition to the requirement to explain the strength of the \ion{N}{iv}] line, an important constraint is to not over-predict other emission lines within the FORS2 spectrum. 
The absence of \ion{N}{v} 1243~\AA\ has already been used to argue against an AGN-like ionizing continuum but it also limits the temperature of the blackbody to $\lesssim 100$kK. 
The \ion{O}{iv}/\ion{Si}{iv} complex from 1394--1413~\AA\ increases in strength at the higher temperatures and is still predicted to be significant in our fiducial model (see Table~\ref{em_lines}). 
It is not seen in our data but, since we have no a priori information about the abundance of silicon, we do not consider this to be sufficient grounds for rejecting the model. The \ion{C}{iv}\ resonance doublet is very close to the long wavelength limit of the observed spectrum and its detection is dependent on the effective redshift which can be influenced, like \lya, by absorption. If it has the same redshift as \ion{N}{iv}], the observational limit is below the prediction of our fiducial model by a factor of nearly 20 while if it is at the redshift of \lya\ it would be undetected \citep[as discussed by][]{van09b}.

\subsection{Sample modelling}

The source  \obj\  is the only member of our sample that contains a clearly detected second emission line in the FORS2 spectroscopy. There is, however, 
one source with groundbased IR spectroscopy that allows the measurement of [\ion{O}{ii}], [\ion{O}{iii}] and \hbeta\  line strengths. This is object CDFS~6664 (GDS~J033233.33$-$275007.4) at
a redshift of 3.791
from the AMAZE programme \citep{mai08}.

The presence of significant [\ion{O}{ii}] emission (i.e., comparable to \hbeta) rules out the high temperature and high ionization parameter models that we need to explain strong \ion{N}{iv}] 
emission. Since all of the spectra reported by \citet{mai08}\ do show [\ion{O}{ii}], their use of the locally calibrated R$_{23}$\ parameter and other line ratios to estimate metallicity is 
justified although it would not be for a \ion{N}{iv}]-emitter, since it was not designed for the high ionization parameter cases.
While we might expect the higher redshift sources to show evidence of hotter ionizing stars, the Lynx arc is at a similar redshift 
to the AMAZE sources but shows no [\ion{O}{ii}] and strong [\ion{O}{iii}] and \ion{N}{iv}] lines. From Figures~\ref{sed1} and \ref{sed2}, we see that most of our sample exhibit the rather flat 
spectra in  f$_\nu$ that is a characteristic of the nebular models. However, without the evidence provided by emission line measurements in addition to \lya, it is not possible for us to 
disentangle the stellar and nebular contributions. The relative importance of starlight and nebular emission in high redshift starforming galaxies has been discussed  by \citet{zac08} who 
model the effect of individual strong emission lines on multi-band SED observations. As we have confirmed, both with the Lynx arc and with  \obj\, the [\ion{O}{iii}] lines 
can carry enough flux to affect significantly the magnitudes in even rather broad photometric bands and it is in the regime where its strength is proportional to oxygen abundance. The presence 
of a significant excess flux in the band that includes these lines can be a hint that we are dealing with an SED with a significant nebular component.

\section{Discussion}

The model we employ for  \obj\  is clearly rather rudimentary and is not expected to produce a detailed match to even the somewhat modest data set that we have. Simply summing models 
with a range of model parameters, notably density, would be more realistic and would tend to produce a better match to the observed SED at the longer wavelengths. However, the general properties of this kind of model are clear: the 
nebular continuum is approximately flat in f$_\nu$  and is modulated by the presence of strong lines in particular photometric bands. 

Although there appear to be no low redshift analogues of \obj, the recent discovery of ``Green Peas'' by the Galaxy Zoo project \citep{car09} does, however, show that there exists a population of compact, strong [\ion{O}{iii}]  emitting sources with high specific star formation rates, low metallicities and low reddening in the local universe. These sources demonstrate that vigorously star forming galaxies can exhibit SEDs with very significant nebular contributions to broad-band photometric measurements.

Arguments for thermal rather than non-thermal (AGN) photoionization were given in section 3.1. It should also be noted that the upper limit on the MIPS 24~$\mu$m flux (Table~\ref{photometry}\ and Figure~\ref{sed_1}) excludes the steeply-rising SED from warm dust typical of AGN of types 1 and 2 \citep{mil08}. Upper limits (3$\sigma$) are available also on the MIPS 70 and 160~$\mu$m fluxes of 2,500 and 33,000~$\mu$Jy respectively (Dickinson, M., priv. comm.; \citet{cop09}) but these values are, unfortunately, not strong discriminants. The source DLS~1053$-$0528 at $z=4.02$ discussed by \citet{gli07} appears in a list of low luminosity type~1 AGN because its emission lines have a broad component with a FWHM $ > 1000$~km s$^{-1}$.  However, aside from this characteristic, this object shares emission line and SED characteristics more closely with \obj\ and the Lynx arc than with typical AGN, notably the great relative strength of the nitrogen line with  \ion{N}{iv}]/ \ion{C}{iv} $> 1$.

The lower limit on the baryonic mass for our fiducial model is the sum of the masses of the ionizing stars plus the mass of the ionized nebula. To estimate the first of these components we use 
the zero metallicity models of \citet{sch02} which give a value of $Q$(H) per star. We choose his $T_{\rm eff} = 85$~kK model since it produces a similar ionization structure to our fiducial model. 
Such stars have a mass of 50~M$_{\odot}$ and $Q({\rm H}) = \sim 3\times 10^{49}$ photons~s$^{-1}$. Given the $Q$(H) required for our model of $ 3\times 10^{55}$ photons~s$^{-1}$, we need a million stars 
with a total mass of $5\times 10^7~{\rm M}_{\odot}$. The mass of ionized gas from the fiducial CLOUDY model is $2.7 \times 10^8~{\rm M}_{\odot}$, giving a total minimum mass of 
$3.2 \times 10^8~{\rm M}_{\odot}$, more than two orders of magnitude smaller than the stellar masses estimated by \citet{sta07} and \citet{wik08}.

The need to properly include the nebular emission in high redshift starforming galaxies is not a new idea, e.g., \citet{sch02,zac08}, but we believe that the Lynx arc at redshift 3.4 (F03) and 
 \obj\  at $z=5.6$\ are the only such objects known where the entire observed SED longward of \lya\  may be contributed by nebular emission. The implication of this is that most 
of the stellar flux is emitted in the Lyman continuum. As shown by F03, a stellar population with a normal IMF that produces sufficient ionizing photons to produce the observed emission line 
fluxes would be easily detected in the restframe UV. Unless we have a mechanism, such as a dust screen, that selectively absorbs this starlight in our line of sight, we are forced to the conclusion 
that we have a large population of stars with an effective temperature close to 100~kK. The absence of light from cooler stars implies that the cluster is either very young or has an intrinsically top-heavy IMF. The latter is an expectation for Population~III stars. While we have no direct method to 
measure the {\em stellar} metallicity in these two objects, we do have estimates of the {\em nebular} metallicity of $\approx5$\% solar. This may seem high compared with normal estimates for a 
Pop~III environment \citep{bro01} but we need to better understand the rate of self-pollution within a cluster of a million stars in a deep potential well with temperatures around 80--100~kK and lifetimes of just a few 
million years.

Our modelling has been carried out without the inclusion of dust reddening. It is possible, however, to maintain a reasonable fit with the same model parameters other than an increased intrinsic 
luminosity (and total mass) by including $E(B-V)\lesssim0.1$ of Galactic dust for a screen model or $E(B-V)\lesssim0.2$ for a mixed model. The presence of some dust may be necessary to explain 
the \lya\ profile \citep{van09b}.

Once the need for a hot, thermal photoionization source is identified, it is natural to associate this solely with 
the presence of massive stars. Other sources of thermal radiation could be 
present, e.g., from shocks associated with a high supernova rate but, even if the entire kinetic luminosity of a core-collapse supernova $\sim 10^{51}$~erg \citep{woo86} could be converted to Lyman continuum photons, this falls far short of the $\sim 10^{54}$~erg of ionizing radiation emitted by a massive star during its lifetime \citep{sch02}.

\section{Conclusions}

The possibility of starforming galaxies at high redshift, where very hot ionizing stars are expected, having restframe UV through MIR SEDs containing significant nebular contributions has been 
discussed for the last decade. The detection of emission lines in addition to \lya\ provides, by using photoionization models, a way of quantifying this contribution. Compact objects containing very 
hot ionizing stars capable of producing lines such as \ion{N}{iv}] and \ion{He}{ii} will also emit strong nebular continuum which we show can dominate the Rayleigh-Jeans tail from the ionizing 
stars. The presence of other strong lines falling within a photometric band can result in SED features than can be confused with jumps such as the Balmer break in conventional stellar population 
models. It is important to realise that the detection of a luminous emission line such as \ion{N}{iv}] {\em necessarily implies} a substantial nebular contribution to the overall SED and therefore effectively excludes models which use purely stellar continuum contributions. The large equivalent widths of the emission lines can be used to limit the contribution of stars significantly cooler than 100~kK to the SED and so constrain the age of the cluster or its stellar IMF.

These highly ionized nebul\ae\  appear very different from the majority of \ion{H}{ii} regions in the local universe due to the higher temperature stars and higher ionization parameters. The restframe UV spectrum is 
rich with emission lines such as the intercombination doublets of C, N, O and Si which can be used for reliable abundance determinations. 

Determined efforts with large groundbased telescopes 
equipped with efficient red and NIR spectrographs may allow us to observe some of the first stellar nucleosynthesis products from the young universe.

\begin{acknowledgements}
We thank Mario Nonino for so quickly finding this source in the GOODS spectroscopy after he had been tipped off about the importance of the \ion{N}{iv}] line.
Daniel Schaerer and Daniel Pequinot gave us much useful modelling advice. We thank Francesca Matteucci for helpful discussions about the chemical evolution of galaxies. AR would like to thank 
Massimo Stiavelli for hosting her visit to STScI to learn and discuss matters pertaining to early star formation. In particular, we thank Eros Vanzella for his sterling work on the 
reduction and analysis of the GOODS/FORS2 spectroscopy and both he and Piero Rosati for many inputs to and discussions about this work.
\end{acknowledgements}

\begin{landscape}
\begin{table}
\caption{Photometry ($AB$~magnitudes) of sample sources in the GOODS passbands with their 1$\sigma$ errors.}
\label{photometry} 
\centering 
\tiny
\begin{tabular}{llllllllllllll} 
\hline\hline 
GDS-id   &  $r_{e}$(")   &  $z$   &  IRAC4   &  IRAC3   &  IRAC2   &  IRAC1   &  $Ks$   &  $H$   &  $J$   &  $z$   &  $i$   & $V$  &  $B$  \\
\hline
J033229.14-274852.6   & 0.138 & 3.597 & -24.54 & -24.40 & 24.15$\pm$0.21 & 24.32$\pm$0.15 & 24.20$\pm$0.10 & 24.56$\pm$0.13 & 24.60$\pm$0.13 & 24.40$\pm$0.03 & 24.48$\pm$0.03 & 24.89$\pm$0.02 & 26.25$\pm$0.09\\
J033201.84-274206.6   & 0.169 & 3.603 &  ...   &  ...   & 23.32$\pm$0.14 & 23.59$\pm$0.11 & 23.77$\pm$0.08 & 25.21$\pm$0.18 & -25.90 & 25.00$\pm$0.06 & 25.06$\pm$0.05 & 25.14$\pm$0.03 & 27.16$\pm$0.37\\
J033217.13-274217.8   & 0.127 & 3.617 & -24.50 & -24.50 & 24.80$\pm$0.27 & 25.04$\pm$0.22 & 25.04$\pm$0.15 & 24.67$\pm$0.13 & 25.72$\pm$0.22 & 24.98$\pm$0.05 & 25.03$\pm$0.04 & 25.12$\pm$0.03 & 26.92$\pm$0.17\\
J033233.33-275007.4   & 0.138 & 3.791 & -24.44 & -24.96 & 25.03$\pm$0.30 & 25.21$\pm$0.23 & 24.99$\pm$0.15 & 24.81$\pm$0.14 & 24.62$\pm$0.13 & 24.70$\pm$0.04 & 24.72$\pm$0.03 & 24.99$\pm$0.02 & 27.35$\pm$0.12\\
J033236.83-274558.0   & 0.153 & 3.797 & -23.97 & -24.57 & 24.11$\pm$0.24 & 24.10$\pm$0.16 & 24.55$\pm$0.13 & 24.57$\pm$0.13 & 24.55$\pm$0.13 & 24.54$\pm$0.02 & 24.62$\pm$0.02 & 25.16$\pm$0.02 & 27.50$\pm$0.14\\
J033240.38-274431.0   & 0.111 & 4.12 & -24.48 & -24.82 & 25.10$\pm$0.37 & 24.59$\pm$0.21 & 24.93$\pm$0.15 & 24.59$\pm$0.13 & 25.38$\pm$0.19 & 24.99$\pm$0.04 & 25.07$\pm$0.03 & 25.64$\pm$0.04 & -28.70\\
J033234.36-274855.8   & 0.251 & 4.138 & -23.91 & 23.70$\pm$0.40 & 23.47$\pm$0.15 & 23.34$\pm$0.10 & 23.63$\pm$0.08 & 24.03$\pm$0.10 & 24.29$\pm$0.11 & 24.10$\pm$0.02 & 24.25$\pm$0.02 & 25.23$\pm$0.03 & 28.00$\pm$0.29\\
J033248.24-275136.9   & 0.146 & 4.374 & -24.54 & -24.54 & 25.00$\pm$0.31 & 24.46$\pm$0.14 & 24.93$\pm$0.15 & 25.82$\pm$0.25 & 24.75$\pm$0.13 & 24.81$\pm$0.04 & 24.91$\pm$0.03 & 25.78$\pm$0.05 & -28.51\\
J033257.17-275145.0   & 0.173 & 4.76 & -24.55 & -24.54 & 24.87$\pm$0.25 & 24.32$\pm$0.13 & 24.36$\pm$0.11 & 24.77$\pm$0.14 & 25.22$\pm$0.17 & 24.54$\pm$0.03 & 24.53$\pm$0.03 & 25.79$\pm$0.06 & -28.60\\
J033229.29-275619.5   & 0.156 & 4.762 & 21.31$\pm$0.24 & 22.52$\pm$0.25 & 23.05$\pm$0.12 & 23.00$\pm$0.08 & ...   &  ... &  ...  & 24.86$\pm$0.05 & 25.12$\pm$0.04 & 26.93$\pm$0.14 & -28.86\\
J033210.03-274132.7   & 0.112 & 4.811 & -24.5 & -24.60 & 24.46$\pm$0.31 & 24.27$\pm$0.16 & 25.25$\pm$0.18 & 24.54$\pm$0.13 & 25.02$\pm$0.19 & 24.94$\pm$0.05 & 25.12$\pm$0.05 & 26.78$\pm$0.12 & -28.61\\
J033221.30-274051.2   & 0.155 & 5.292 & -24.5 & -24.29 & 25.15$\pm$0.32 & 25.68$\pm$0.29 & 25.32$\pm$0.18 &  ...   & 25.71$\pm$0.22 & 25.14$\pm$0.08 & 25.13$\pm$0.10 & 28.33$\pm$0.25 & -29.50\\
J033245.43-275438.5   & 0.190 & 5.375 & -24.5 & -24.50 & 24.02$\pm$0.19 & 24.71$\pm$0.18 & 24.86$\pm$0.14 & 24.94$\pm$0.15 & 24.88$\pm$0.14 & 25.00$\pm$0.05 & 25.70$\pm$0.07 & -28.97 & -28.76\\
J033237.63-275022.4   & 0.263  & 5.518 & -24.24 & -24.24 & 24.57$\pm$0.30 & 24.69$\pm$0.24 & 25.40$\pm$0.31 & 24.60$\pm$0.20 & 25.82$\pm$0.25 & 25.71$\pm$0.10 & 26.28$\pm$0.22 & -29.50 & -29.50\\
J033218.92-275302.7   & 0.109 & 5.563 & 23.15$\pm$0.20 & 23.70$\pm$0.31 & 23.37$\pm$0.08 & 23.20$\pm$0.07 & 24.45$\pm$0.11 & 24.96$\pm$0.14 & 24.73$\pm$0.11 & 24.53$\pm$0.04 & 25.18$\pm$0.05 & 27.48$\pm$0.03 & -28.75\\
J033225.61-275548.7   & 0.110 & 5.786 & -24.61 & -24.60 & 24.45$\pm$0.19 & 24.04$\pm$0.11 & 25.04$\pm$0.15 & 24.81$\pm$0.14 & 24.61$\pm$0.12 & 24.42$\pm$0.04 & 26.32$\pm$0.17 & 27.67$\pm$0.23 & -29.51\\
J033246.04-274929.7   & 0.120 & 5.787 & -24.49 & -24.49 & 25.65$\pm$0.39 & 26.10$\pm$0.32 & 25.49$\pm$0.19 & 25.66$\pm$0.21 & 25.85$\pm$0.22 & 26.26$\pm$0.14 & 27.83$\pm$0.45 & -29.08 & -29.21\\
J033240.01-274815.0   & 0.120 & 5.828 & -24.6 & -24.80 & 24.48$\pm$0.20 & 24.49$\pm$0.14 & 25.32$\pm$0.17 & 25.00$\pm$0.13 & 25.07$\pm$0.15 & 25.21$\pm$0.06 & 26.52$\pm$0.15 & -29.65 & -29.46\\
\hline 

\end{tabular}
\begin{list}{}{}
\item[] Negative numbers indicate upper limits. Also given are the half-light radii and the 
redshift measured from \lya.
\end{list}
\end{table}
\end{landscape}

\bibliographystyle{aa}
\bibliography{ref}

\end{document}